\documentstyle[aps]{revtex}

\begin{document}
\flushbottom
\def\thepage{\roman{page}}
\title{\vspace*{1.5in}
On generation of metric perturbations during preheating}
\author{P. Ivanov}

\address{Theoretical Astrophysics Center, Juliane Maries Vej
30, 2100 Copenhagen, {\O} Denmark\\
Astro Space Center of P. N. Lebedev Institute, Profsoyznaya 84/32,
117810 Moscow, Russia}
\maketitle

\input{epsf}

\newcommand{\be}{\begin{equation}}
\newcommand{\ee}{\end{equation}} \newcommand{\g}{\nabla}
\newcommand{\de}{\partial}    \newcommand{\ha}{\frac{1}{2}}
\newcommand{\ci}[1]{\cite{#1}}  \newcommand{\bi}[1]{\bibitem{#1}}
\newcommand{\noi}{\noindent}

\newcommand{\ga}{\alpha}
\newcommand{\gb}{\beta}
\newcommand{\gc}{\gamma}
\newcommand{\gd}{\delta}
\newcommand{\gep}{\epsilon}
\newcommand{\gee}{\varepsilon}
\newcommand{\gz}{\zeta}
\newcommand{\get}{\eta}
\newcommand{\gth}{\theta}
\newcommand{\gthh}{\vartheta}
\newcommand{\gi}{\iota}
\newcommand{\gk}{\kappa}
\newcommand{\gl}{\lambda}
\newcommand{\gm}{\mu}
\newcommand{\gn}{\nu}
\newcommand{\gks}{\xi}
\newcommand{\go}{\0}
\newcommand{\gp}{\pi}
\newcommand{\gpp}{\varpi}
\newcommand{\gr}{\rho}
\newcommand{\grr}{\varrho}
\newcommand{\gs}{\sigma}
\newcommand{\gss}{\varsigma}
\newcommand{\gt}{\tau}
\newcommand{\gu}{\upsilon}
\newcommand{\gf}{\varphi}
\newcommand{\gff}{\varphi}
\newcommand{\gx}{\chi}
\newcommand{\gps}{\psi}
\newcommand{\gw}{\omega}
\newcommand{\gG}{\Gamma}
\newcommand{\gD}{\Delta}
\newcommand{\gTh}{\Theta}
\newcommand{\gL}{\Lambda}
\newcommand{\gKs}{\Xi}
\newcommand{\gP}{\Pi}
\newcommand{\gS}{\Sigma}
\newcommand{\gU}{\Upsilon}
\newcommand{\gF}{\phi}
\newcommand{\gPs}{\Psi}
\newcommand{\gW}{\Omega}

\newcommand{\ti}{\tilde}
\newcommand{\Li}{{\cal L}}
\newcommand{\ra}{\rightarrow}
\newcommand{\pa}{\partial}
\newcommand{\ov}{\overline}
\newcommand{\fad}{\frac{\Delta T}{T}}
\newcommand{\lan}{\langle}
\newcommand{\ran}{\rangle}

\begin{abstract}
We consider the generation of the scalar mode of the metric perturbations 
during preheating stage in a two field model with the potential $V(\phi, \chi)=
{m^{2}\phi^{2}\over 2}+{g^{2}\phi^{2}\chi^{2}\over 2}$. We discuss two possible
sources of such perturbations: a) due to the
coupling between the perturbation
of the matter field $\delta \chi$ and the background part of the matter field
$\chi_{0}(t)$, b) due to non-linear
fluctuations in a condensate of ``particles'' of
the field $\chi$. Both types
of the metric perturbations 
are assumed to be small, and
estimated using the linear theory of the metric perturbations.
We estimate analytically the upper limit of the amplitude
of the metric perturbations for all scales in the limit of so-called broad
resonance, and show that
the large scale
metric perturbations are very small, and taking them into account does 
not influence the standard picture of the production of the metric 
perturbations in inflationary scenario.

PACS number(s): 98.80.Cq
\end{abstract}

\section{Introduction}

In the modern scenario of creation of the matter in the Universe
there is a stage called preheating, when the matter (usually
imitating by
a massless scalar field) is generating from vacuum fluctuations
due to the effect of parametric resonance [1,2,3]. 
Namely it is assumed that
the matter field $\chi$ is coupled with inflaton field $\phi$, and the
coupling constant $g$ is large. During inflation and right after
that period the matter field has a large effective ``mass''
determined by this coupling, and each mode of the matter
field with sufficiently small wavenumber
evolves by a standard manner,  oscillating with
a frequency proportional to this ''mass'' and adiabatically
decreasing amplitude. The inflaton starts to oscillate itself 
after inflation,
and near the moments of time when $\phi(t)\approx 0$ the matter field
effectively loses its ``mass'', the adiabatic approximation breaks and
a possibility of resonant growth of the matter field amplitude
and the corresponding occupation numbers of $\chi$
``particles'' appears. In fact, this effect may lead to exponential growth
of the 
occupation
numbers, and the resulting distribution of $\chi$ ``particles'' is
strongly nonthermal. After some moment of time $t_{*}$ (end of the first
stage of preheating) the energy density of the $\chi$ ``particles'' becomes
comparable with the energy density of the inflaton, and back reaction
processes  become influence the dynamics of inflaton field $\phi$ and
the Universe. It is believed that during some short
period of time $\delta
t$ after $t_{*}$ the
$\chi$ particles are thermalized by rescattering effects, and 
after thermalization the energy density of
these ``particles'' evolves according to the standard picture of the Hot
Big Bang model. Such scenario of the first stage of preheating is
called the scenario of broad resonance [2,3].

As was mentioned by a number of authors (e.g [4]), 
the characteristic value of the growth
rate of the matter field modes does not depend significantly on the 
value of wavenumber $k$ for the modes with sufficiently small $k$.
At first glance this 
fact gives a very interesting possibility of amplification of the
matter field modes 
and metric perturbations at exponentially large scales,
say, corresponding to the
scale of the present horizon. 
In this paper we consider the simplest model of
preheating with two scalar fields, the inflaton field
$\phi$ and the matter field
$\chi$ and the potential 
$V(\phi, \chi)={m^{2}\phi^{2}\over 2}+{g^{2}\phi^{2}\chi^{2}\over 2}$,
and show that in this model the
value of metric perturbations at the scale of present horizon
is suppressed by an extremely small factor,
and therefore taking into account the stage of preheating
does not lead to
any modifications of the standard picture of 
generation of cosmological perturbations in inflationary scenario.
We have two basic arguments supporting this conclusion. At first,
the field $\chi$ has a large effective mass before the stage of 
preheating
(much larger than the Hubble parameter at the end of inflation),
and as a consequence the spectrum of initial field fluctuations 
$\delta \chi_{k}$
is strongly
suppressed at small $k$ at the time $t_{in}$ of the beginning of the
preheating stage, the rms value of the field fluctuations
$\delta \chi_{rms}\propto k^{3/2}$ (also [5]). 
During preheating the field
modes may grow exponentially fast, but the characteristic growth rate
$G={{\dot {\delta \chi_{k}}} \over {\delta \chi_{k}}}$
is constrained by some maximal value $\mu_{max}m$, where numerical
constant 
$\mu_{max}={1\over \pi}\ln{(1+\sqrt 2)}\approx 0.28$ [3]. 
The first stage of preheating ends at the time $t_{*}\approx 50\div
100m^{-1}$,
and this estimate does not depend strongly on the 
parameters of the theory 
(see [3], and also the eqns. $(22)$, $(23)$ below).
Thus, at the end of the first stage of preheating, 
the r.m.s value of the
field fluctuations contains a factor 
$$\delta \chi_{rms}(t_{*})< G\delta \chi_{rms}(t_{in})
\sim  e^{-3N/2+\mu_{max}m t_{*}}
= e^{-47}e^{-1.5(N-50)+0.28(mt_{*}-100)}, \eqno (1)$$
where the $N$ is the number of e-folds, and we choose the standard
value $N=50$ to represent the scale of present horizon.
Here we express the amplitude of the field perturbation in terms
of the natural Plank units. 

Secondly,
the suppression
factor for the metric perturbations
may be even much smaller than the estimate $(1)$ if one uses
the standard theory. 
In this theory the contribution of the second field
$\chi$ in the scalar mode of metric perturbations
is determined by the terms containing multiplication of
homogeneous ``background'' part of the field $\chi_{0}(t)$ and the
perturbed
part $\delta \chi$, and their derivatives (see eqns. (25,26) below). 
The amplitude of homogeneous part of the field behaves like the mode
$\delta \chi_{k}$ with wavenumber $k=0$, and
is constrained by the
fact that the field $\chi_{0}$ cannot contribute to the total
energy density during the last stage of inflation. Assuming that  
during last $N$ e-folds the dynamics of the Universe
has been controlled by     
the inflaton field $\phi$, after the end of the first stage of 
preheating the amplitude $\chi_{0}$ should contain the same factor
$(1)$. Therefore the rms amplitude of the metric perturbations
$\delta_{rms}$ contains the factor $(1)$ squared:
$$\delta_{rms}(t_{*})
< e^{-94}e^{-3(N-50)+0.56(mt_{*}-100)}, \eqno 2$$
and is suppressed by enormously small factor $e^{-94}$ for
the typical parameters of the theory. This estimate is however 
changed in a more self-consistent
approach to the evolution of the perturbations
during preheating. In the theory of preheating the role of background
is effectively played by a condensate of $\chi$ particles with relatively 
large
wavenumbers ( a typical wavenumber of such particles $k_{crit}$
is always
larger than the characteristic wavenumber corresponding to the 
scale of cosmological horizon during the 
preheating stage, see [3] and also
the eq. $(13)$). The fluctuations of energy-momentum tensor
of such condensate can give rise to additional metric perturbations,
which are second order with respect to the perturbations of the
matter field, and therefore cannot be obtained in the frameworks of 
the standard perturbation theory. The estimate shows that 
the fluctuations of the energy-momentum tensor decrease with scale
proportional to $k^{3/2}$ (similar to the rms value of $\delta \chi$
field, see eq. $(47)$), 
and is of order of unity at the scale corresponding to the
critical wavenumber $k_{crit}$ at the time $t_{*}$. Since in
the large scale limit the metric fluctuations are of the order
of the energy density fluctuations, the estimate
$\delta_{rms}(t_{*})\sim \delta \chi_{rms}(t_{*})$ 
is a more reliable estimate in the more realistic approach to
the dynamics of preheating, and this estimate still contains a very
small number $e^{-47}$ for the scale corresponding to the present horizon. 
In fact, it can be shown (see Section 3) that
the metric perturbations of this kind take their maximal value at the scale
corresponding to the horizon size at the time of the 
end of preheating, and this value is smaller
than $\sim 10^{-3}$ for the broad resonance case
\footnote{Note, that when estimating the metric perturbations
induced by the non-linear terms, we do not take into account the
contribution of such fluctuations in the dynamics of the background
model. This contribution is of order of the leading term at the end of
preheating, and can change our estimates on a numerical factor.
We believe that this factor is of order of unity, and cannot change
our results significantly.}.
Therefore in the simplest models
the transition from inflation to
the hot Big Bang proceeds smoothly, without significant 
generation of the metric perturbations.

We present our arguments in a more rigorous form
in the next two Sections. In Section $2$ we introduce the
basic ideas 
of preheating theory, and estimate the time $t_{*}$ of duration
of the first stage of preheating.
In Section $3$ we discuss the application of the theory of cosmological
perturbations to our case of two interacting fields, and estimate the
upper limit on the metric perturbations.

We use below the natural system of units, and set the Plank mass 
$M_{pl}=\sqrt{8\pi}$.

\section{Preheating in the regime of broad resonance}      
  
The theory of initial stage of preheating has been developed in the
paper [3], and we will closely follow this paper. For our purposes
we need to know the evolution of both fields $\phi$ and $\chi$,
and also the evolution of the scale factor $a(t)$. As usually we divide
both fields on background parts and perturbations 
$\phi=\phi_{0}(t)+\delta \phi(t, \vec x), 
\chi=\chi_{0}(t)+\delta \chi(t, \vec x)$, and apply Fourier
transform to the perturbed parts. We temporary neglect the influence
of the metric perturbations in this Section, and consider
this effect later on.
Assuming that the contribution of
the field $\chi$ in the energy density and the pressure
is negligible, we can use 
the standard expression describing the evolution of the scale factor and
the field $\phi_{0}$ in the theory of massive scalar field (see
e.g. [6]):
$$a(t)\approx a_{0}\tau^{2/3}, \eqno 3$$
$$\phi_{0}(t)\approx \phi_{in}{\sin \tau \over \tau}, \eqno 4$$  
where $\phi_{in}=2\sqrt{{2\over 3}}$ is a characteristic value of the
field in the beginning of preheating stage, $a_{0}$ is an 
``initial'' value
of the scale factor, dimensionless time $\tau =mt$, and the preheating 
stage begins when approximately $\tau =\tau_{in}\approx 1$.

The evolution equation for the perturbed part of the field $\chi$ can
be conveniently written in the form:
$$\ddot X_{k}+\omega_{k}^{2}X_{k}=0, \eqno 5$$
where $X_{k}$ is the rescaled field amplitude: 
$X_{k}=a^{3/2}\delta \chi_{k}$, $\omega_{k}$ is the effective frequency:
$$\omega_{k}^{2}={k^{2}\over a^{2}}+g^{2}\phi_{in}^{2}{({\sin \tau
\over \tau})}^{2} +\Delta, \eqno 6$$
and the correction $\Delta=-({3\over 4}H^{2}+{3\over 2}{\ddot a\over a})$.
Hereafter $H={\dot a \over a}$ is the expansion rate. 
The positive frequency solutions of this equation $X_{+}$
determine a vacuum state,
and must have a form:
$$X_{+}={1\over {(2\pi)}^{3/2}\sqrt{2\omega_{k}}}e^{-i\theta} \eqno 7$$
at the moments of time sufficiently close to $\tau_{in}$
\footnote{A care should be taken when specifying the positive frequency
solution for the modes with wavelengths larger than the
horizon scale.
Strictly speaking the vacuum state should be the standard Bunch-Davies 
vacuum state for a massive field, but a more accurate expression give
essentially the same result.}
Here
$\theta=\int^{t} dt^{'} \omega $
\footnote{Wherever it is possible we will drop
the index $k$ in our expressions.}.
The solution
of the equation $(5)$
can also be represented
in another form by introducing two complex functions
$\alpha, \beta$, which are constrained by the normalization condition
${|\alpha|}^{2}-{|\beta|}^{2}=1$:
$$X_{+}={1\over {(2\pi)}^{3/2}}
({\alpha (t)\over \sqrt{2\omega}}e^{-i\theta}+
{\beta (t)\over \sqrt{2\omega}}e^{i\theta}). \eqno 8$$
To reconcile this representation with $(7)$, we should set 
$\alpha(\tau\sim \tau_{in})= 1$, and
$\beta(\tau\sim \tau_{in})= 0$.
Obviously, the representation $(8)$
is specially convenient if the solution of eq. $(5)$ is close to its
adiabatic approximation, and $\alpha, \beta \approx const$.  
For the general case, the evolution of the functions $\alpha$, 
$\beta$ follows from the eq. $(5)$:
$$\dot \alpha ={\dot \omega \over 2\omega}e^{2i\theta}\beta,\quad
\dot \beta ={\dot \omega \over 2\omega}e^{-2i\theta}\alpha. \eqno 9$$
If the adiabatic condition ${\dot \omega \over \omega^{2}}\ll 1$ 
is satisfied, the functions $\alpha$, $\beta$ are approximately constant,
and no additional ``particles'' of the field $\chi$ are produced, the field
oscillates with the frequency $\approx \omega$, and the field
amplitude decays
as $a^{-3/2}\sim \tau^{-1}$. The adiabatic approximation
breaks when the field $\phi$ is close to zero, and the time $\tau$
is close to $\tau_{j}=\pi j$ (the integer index $j$ must be much larger
than unity for the validity of the approximate equations $(3,4)$). Rewriting
the adiabatic condition near the points $\tau=\tau_{j}$, we have
$${\dot \omega \over \omega^{2}}\approx {m\tau_{j} \over g\phi_{in} 
{\Delta \tau}^{2}}={1\over 2q{(\tau_{j})}^{1/2}{\Delta \tau }^{2}}, \eqno 10$$
where $\Delta \tau =\tau -\tau_{j}$, and the parameter $q(\tau)$ 
characterizes the strength of the resonance [3]:
$$q={2\over 3}{({g\over m})}^{2}{1\over \tau^{2}}. \eqno 11$$
We will also use $q_{0}=q(\tau =1)$ to parameterize our expressions.
The regime of broad resonance corresponds to very large values of
$q_{0}$, see below. 

Thus 
when the parameter $q$ is very large, the functions $\alpha$ and $\beta$
are approximately constant between the moments of time $\tau_{j}$, but 
when the time is very close to $\tau_{j}$, 
the functions $\alpha$ and $\beta$ can be changed. The rule of
change can be written as an iterative
mapping between the functions $\alpha_{j-1}$, $\beta_{j-1}$ and
$\alpha_{j}$, $\beta_{j}$ corresponding to the time periods $\tau_{j-1} <
\tau < \tau_{j}$ and $\tau_{j} < \tau < \tau_{j+1}$ [3]:
$$\left( \matrix{\alpha^{j} \cr \beta^{j}} \right )=\left ( \matrix {
a & b \cr b^{*} & a^{*}}\right )=\left (\matrix {\alpha^{j-1}\cr \beta^{j-1}}
\right ), \eqno 12$$
where
$a=\sqrt{1+e^{-\pi \kappa^{2}}}e^{i\phi},$
and 
$b=ie^{-{\pi \over 2}\kappa^{2}+2i\theta^{j}},$
and $\theta^{j}=\int^{t_{j}}dt \omega$, $t_{j}=\tau_{j}/m$, 
$\phi=arg \Gamma({1+i \kappa^{2}\over 2})+
{\kappa^{2}\over 2}(1+2\ln{2\over\kappa^{2}})$, and $\kappa=k/k_{*}$,
$k_{*}$ is a characteristic value of comoving momentum:
$k_{*}=2^{1/2}a(t)q^{1/4}m$. If the wavenumber is much larger than
a critical cutoff wavenumber:
$$k_{crit}={k_{*}\over \pi}=\sqrt{2\over \pi}a(t)q^{1/4}m=
\sqrt{{2\over \pi}}a_{0}q_{0}^{1/4}\tau^{1/6}m, \eqno 13$$  
the effect of particle creation is exponentially damped.
It is important to note that the bulk of the produced particles has 
values of wavenumbers close to the cutoff value $(13)$, 
the corresponding wavelengths of these particles are well inside
the cosmological horizon. The particles occupation numbers 
$n_{j}={|\beta_{j}|}^{2}$ can also be expressed by an iterative way:
$$n_{j}\approx (1+2e^{-\pi\kappa^{2}}-2\sin {\theta_{tot}}
e^{-{\pi \over 2}}\kappa^{2}\sqrt {1+e^{-\pi \kappa^{2}}})n_{j-1}, \eqno 14$$
where $\theta_{tot}=2\theta^{j}-\phi +arg (\beta^{j-1})-arg (\alpha^{j-1})$,
and the limit of large occupation numbers $n_{j-1} \gg 1$ is assumed
hereafter.
In particular, from the eq. $(14)$ 
it follows that the number of ``particles'' can
either increase or decrease depending on the value of $\theta_{tot}$.
The evolution of $\theta_{tot}$ is rather complicated in the limit
of large $g$, but for our purposes 
it is sufficient to note that the occupation number
 cannot increase more than $3+2\sqrt 2$ times, and the
amplitude of the field $\sim \sqrt {n_{j}}$ cannot increase more than
$1+\sqrt 2$ times. It is convenient to characterize the change of the
amplitude by the growth rate:
$$\mu_{j}={1\over 2\pi}\ln{{n^{j}\over n^{j-1}}}, \eqno 15$$
and also average the growth rate over the time:
$$\mu^{eff}={\pi \over \tau} \sum_{j} \mu_{j}. \eqno 16$$
From the eqns. $(14-16)$ it follows that neither $\mu_{j}$ nor
$\mu^{eff}$ can exceed the maximal value $\mu_{max}={1\over \pi}
\ln {(1+\sqrt 2)}\approx 0.28$. In fact, the effective growth rate
$\mu^{eff}$ is about two times smaller than its maximal value. The
numerical calculations give $\mu^{eff}$ in the interval $0.1-0.18$
with an average value of order of $0.14$
for the coupling constant $g$ in the interval from $0.9\cdot 10^{-4}$
to $10^{-3}$ [3].    

The background field $\chi_{0}(t)$ obeys the same equation $(5)$ provided
the wavenumber $k$ is set to zero. Therefore we can obtain the expression
for the evolution of the background field combining the positive and 
negative frequency solutions $\delta \chi_{+}$, $\delta \chi_{-}=
\delta \chi_{+}^{*}$. 
Let denote
the value of the field $\chi_{0}$ at the beginning of the preheating stage
as $\chi_{in}$, and its time derivative as $\dot \chi_{in}$. The values of 
these quantities are determined by their evolution at the previous stage of
inflation. Assuming that the contribution of the field $\chi$ in 
the potential has been negligible during last $N$ e-folds, we estimate the
initial amplitude as
$\chi_{in}\approx {m\over g}e^{-3N/2}$, and its time derivative as 
$\dot \chi_{in}\approx \omega (\tau \sim \tau_{in})\chi_{in}
\approx m e^{-3N/2}$. Let introduce a characteristic field amplitude
and angle:
$$\chi_{*}=\sqrt{\chi_{in}^{2}+g^{-2}\dot \chi_{in}^{2}},\quad
\phi_{0}=\arctan {{g\chi_{in}\over \dot \chi_{in}}}. \eqno 17$$
Then 
$$\chi_{0}(t)={{(2\pi a_{0})}^{3/2}\omega_{0}^{1/2}\chi_{*}\over i\sqrt 2}
(e^{i\phi_{0}}\delta \chi_{-} -e^{-i\phi_{0}}\delta \chi_{+})=$$
$$
{\chi_{*}\over 2i}{({a_{0}\over a})}^{3/2}
{({\omega_{0}\over \omega})}^{1/2}
(e^{-i\theta}(\beta^{*}e^{i\phi_{0}}-\alpha e^{-i\phi_{0}})+
e^{i\theta}(\alpha^{*}e^{i\phi_{0}}-\beta e^{-i\phi_{0}})), \eqno 18$$
where the wavenumber $k$ is set to zero in the expressions for
$\alpha$ and $\beta$,
and $\omega_{0}=\omega (\tau=\tau_{in})\sim g$.

We also need the evolution of the field $\delta \phi$, but we discuss it in
the next Section.

Since our order-of-magnitude estimates $(1), (2)$ are exponentially 
sensitive to the value of the moment of time $t_{*}$ of the end of the 
first stage of
preheating, it is very important to obtain a reliable estimate of 
$t_{*}$. Following [3] we assume that the first stage of preheating
ends when
the vacuum expectation value for $\chi^{2}$ gives a contribution to the
potential of order of the leading classical term ${m^{2}\phi^{2}\over 2}$:
$$\left <\chi^{2} \right >(t_{*})={m\over g}. \eqno 19$$
Let estimate the dependence of $\left <\chi^{2}\right >$ on time. 
Using the definition
of the function $\beta$, we have 
$$\left <\chi^{2}\right >(t)={1\over 2\pi^{2}a^{3}\omega}\int^{k_{crit}}_{0}dk 
k^{2}{|\beta|}^{2}, \eqno 20$$
where $k_{crit}$ is given by the eq. $(13)$. In order to calculate 
the integral in the eq. $(20)$ we assume that each mode with a given $k$
grows with its average growth rate $\mu^{eff}\approx 0.14$ when
$k< k_{crit}(t)$, and the growth
rate for the modes with $k > k_{crit}(t)$ is zero. We have:
$$\left <\chi^{2} \right >=
{1\over 12\pi^{3}}{q_{0}^{1/4}m^{2}\over {\mu^{eff}}^{1/2}\tau}
e^{2\mu^{eff}\tau}, \eqno 21$$
where we approximate the value of $\omega$ as $\omega \approx {g\phi_{0}\over
\tau}$. Note that our estimate is slightly different from the estimate
[3] due to a different approximation in the calculation of the integral
in $(20)$. Substituting the eq. $(21)$ in $(19)$, we obtain:
$$\tau_{*}=mt_{*}={1\over 2\mu^{eff}}\ln{(\sqrt{{2\over 3}}{12\pi^{3}
{\mu^{eff}}^{1/2}\tau_{*}\over m^{2}q_{0}^{3/4}})}. \eqno 22$$    
An approximate solution of the eq. $(21)$ can be written as:
$$\tau_{*}\approx \mu^{-1}_{0.14}(107+3.57\ln F), \eqno 23$$
where $F=m^{-2}_{-6}q_{4}^{-3/4}\mu_{0.14}^{1/2}$, and 
$m_{-6}={m\over 10^{-6}}$, $q_{4}={q_{0}\over 10^{4}}$, 
$\mu_{0.14}={\mu^{eff}\over 0.14}$. The time $\tau_{*}$ decreases with
increase of the parameter $q_{0}$, and is bounded by some minimal and
maximal values $\tau_{min}$, $\tau_{max}$. 
For example, if the mass $m$ has its ``canonical'' value
$\sim 10^{-6}$, the parameter $q_{0}$ cannot exceed the maximal value
$q_{max}\sim m^{-2}=10^{12}$, and we have $t_{min}\approx 55$. On the
other hand, the parameter $q(t_{*})$ cannot be smaller than unity for
applicability of our theory. 
\footnote{Of course preheating can go on when
$q(\tau) \ll 1$. 
However, in that case the growth rate $\sim q$ is small, and
preheating is inefficient. In the two fields model
the metric perturbations in the limit of
small $q$ have been discussed in the paper [7]. The similar one field model
has been considered in the paper [8].}
Since $q\sim \tau^{-2}$, we have $q_{min}\sim
10^{4}$ as a minimal possible value of $q_{0}$.
Therefore, the time of the end of the
first stage of preheating is constrained by a condition: $55 < \tau_{*}
< 107$ for $\mu^{eff}\approx 0.14$ and $m \sim 10^{-6}$.  

\section{The metric perturbations during preheating}

Now let us take into account the metric perturbations and consider the
self-consistent theory of perturbations. The shape of the metric 
perturbations depends on gauge. In many applications the gauge-independent
formalism is used, where all perturbed quantities are expressed in terms
of their gauge-independent combinations. These combinations are naturally
linked to some preferable
coordinate system. The most commonly used preferable coordinate systems
are so-called comoving and Newtonian coordinate systems. However, both these
systems are not convenient in our case because the dynamical equations
written in these systems have a singularity when $\dot \phi_{0}=0$. 
Therefore we use a synchronous coordinate system, and to specify 
the gauge we assume that our synchronous system is comoving
at the moment of time $\tau=\tau_{in}$. For the simplicity we also 
neglect the metric perturbations generated during inflation. In a 
synchronous coordinate system the metrics has the following form:
$$ds^{2}=dt^{2}-a^{2}(t)(A\delta^{\alpha}_{\beta}+B^{,\alpha}_{,\beta})
dx_{\alpha}dx^{\beta}, \eqno 24$$
where the Greek indices run from $1$ to $3$, and the summation rule is
used hereafter. The prime stands for differentiation with respect to
$x^{\alpha}$. The perturbed equations of motion are (e.g. [9,10]):
$$H\dot h=\dot \phi_{0}\dot {\delta \phi} +\dot \chi_{0}\dot {\delta \chi}
+{\partial V \over \partial \phi}\delta \phi +{\partial V \over 
\partial \chi}\delta \chi,
\eqno 25$$
$$\dot A=-(\dot \phi_{0}\delta \phi +\dot \chi_{0} \delta \chi), \eqno 26$$
$$\ddot {\delta \phi_{i}} +3H\dot {\delta \phi_{i}}+
{\dot \phi_{i} \dot h \over 2}+{\partial^{2}V\over \partial \phi_{i}
\partial \phi_{j}}\delta \phi_{j}=0, \eqno 27$$
where $h=3A+\Delta B$, $i,j=1,2$, for definiteness index $1$ stands for
the fields $\phi_{0}$, $\delta \phi$, and $2$ for the fields
 $\chi_{0}$, $\delta \chi$. We neglect the terms proportional to
$k^{2}$ in the eqns $(25-27)$, assuming that the wavenumbers of the
perturbations are very small.
The quantities $A$ and $B$ are related
as (e.g. [10]) 
$$ B=\int^{t}{dt^{'}\over a^{3}}\int dt^{``}(aA) \eqno 28$$
As we discussed above
the background field $\chi_{0}$ is very small, and does not
contribute to the expansion law, and in that case 
the equations of motion of the fields perturbations $(27)$ can be
rewritten in a simplified manner
\footnote{Obviously, when approaching the time $t_{*}$
the back reaction of the produced 
$\chi$ ``particles'' influences the motion of our system much more
significantly than the field $\chi_{0}$, and comparable with the contribution
of the field $\phi_{0}(t)$. 
We cannot take into account the back reaction without a very
significant complication of our consideration, but we hope that this
cannot change order-of-magnitude estimates. Therefore, our results should
be treated as semi-qualitative only.}.  
The equation for the field $\delta \chi$
is reduced to the eq. $(5)$, and the equation for the field $\delta \phi$
has a form
$${\delta \ddot \phi}+(3H+{{\dot \phi_{0}}^{2}\over 2H}){\delta \dot \phi}
+({\partial^{2}V\over \partial \phi^{2}}+{\dot {\phi_{0}}\over 2H}{\partial V
\over \partial \phi})\delta \phi =J(\phi, \chi_{0}, \delta \chi), \eqno 29$$
$$J(\phi, \chi_{0}, \delta \chi)=
-({\partial^{2}V\over \partial \phi \partial \chi}
+{\dot {\phi_{0}}\over 2H}{\partial V
\over \partial \chi}\delta \chi+{\dot{\phi_{0}}\dot {\chi_{0}}{\delta
\dot \chi}\over 2H}). \eqno 30$$
In this approximation the fields $\chi_{0}$, $\delta \chi$ enter 
only in the  source term $J(\phi, \chi_{0}, \delta \chi)$, and the formal
solution of the eq. $(29)$ can easily be obtained if the source term is
known as a function of time $J(\phi(t), \chi_{0}(t), \delta \chi (t))= J(t)$:
$$\delta \phi \sim \int^{t}d\xi J(\xi) {\dot \phi_{0}(t)\dot \phi_{0}(\xi)H
\over a^{3}}\int^{t}_{\xi}d\xi^{'}{H\over a^{3}\dot H}, \eqno 31$$
Then one can substitute the solution $(31)$ and the expressions for 
$\chi_{0}(t)$ and $\delta \chi (t)$ into the eqns. $(25)$, $(26)$, and find
the metric perturbations by integration. Unfortunately, this program is
analytically very complicated. Even the numerical solution is not trivial
for all possible parameters of the theory. Still, not everything is lost.
At first we need to estimate only the upper limit on
the metric perturbations, and therefore we can assume that both fields:
the background field $\chi_{0}$ and the perturbation $\delta \chi$ grow
with the maximal possible rate $\mu_{max}m$. In this 
approximation at the time close to $\tau_{*}$
the growth rate is much larger than the expansion rate 
$H={2m\over 3\tau_{*}}\sim
{m\over 100}$, and we can take into account the source term $(30)$ only
during the last Hubble epoch before the end of preheating ${\tau_{*}
-\tau \over \tau_{*}}\ll 1, \tau_{*}\gg 1$. 
Secondly, we should take into account only
the leading terms in the expansion on powers of
the small parameter $q^{-1}(\tau)$.
Using the first approximation, we neglect the terms caused by the
expansion of the Universe in the equation of motion of the field $\delta
\phi$, and write:
$$\delta \ddot \phi +m^{2}\delta \phi =-{\partial^{2} V\over \partial \phi
\partial \chi}\delta \chi
=-2g^{2}\phi_{in}{\sin \tau \over \tau}\chi \delta \chi,
\eqno 32$$
where the term $\chi \delta \chi$ can be written as:
$$\chi \delta \chi ={\chi_{*}\over 2\sqrt 2 i{(2\pi a)}^{3/2}}
{({a_{0}\over a})}^{3/2}{\omega_{0}^{1/2}\over \omega}(s^{j}_{0}+
s^{j}_{-}e^{-2i\theta}+s^{j}_{+}e^{2i\theta}), \eqno 33$$
where 
$$s_{0}^{j}=(2n_{j}+1)e^{i\phi_{0}}-2\alpha_{j}\beta_{j}e^{-i\phi_{0}},
\quad      
s_{+}^{j}=\beta_{j}(\alpha_{j}^{*}e^{i\phi_{0}}-\beta_{j}e^{-i\phi_{0}}),
\quad
s_{-}^{j}=\alpha_{j}(\beta_{j}^{*}e^{i\phi_{0}}-\alpha_{j}e^{-i\phi_{0}}),$$
and
the functions $\alpha_{j}$, $\beta_{j}$ correspond to the period of time
$\tau_{j} < \tau < \tau_{j+1}$. 
In the same approximation we can set $a=a(t_{*})=a_{*}$. 
In this case during the
time period $\tau_{j-1}< \tau < \tau_{j}$ the equation $(32)$ is just
an oscillator equation with a source, represented by a sum of
a constant and an oscillating parts. Assuming $q(t)\gg 1$,
it can be easily seen that
a partial solution determined by the oscillating parts is small, and
the partial solution
of the eq. $(32)$ $\delta \phi_{p}$ is close to a constant:
$$\delta \phi_{p}\approx K(-1)^{j}s_{0}^{j}, \eqno 34$$    
where
$$K={i{(g/m)}^{2}g^{-1/2}\over \sqrt 2}{({a_{0}\over a_{*}})}^{3/2}{\chi_{*}
\over {(2\pi a_{*})}^{3/2}}, \eqno 35$$
and we approximate $\omega_{0}\sim g$. The full solution can be found from
the continuity conditions at the time moments $\tau_{j}$, and has a form:
$$\delta \phi =K({(-1)}^{j}s_{0}^{j}+\sum^{l=j}_{l=0}
c_{l}\cos \tau), \eqno 36$$
where
$$c_{l}=-(s_{0}^{l}+s_{0}^{l-1}).$$
Now we substitute the solutions $(4)$, $(36)$ into
the eq. $(26)$, and by integrating the result
obtain the perturbation
of the scale factor $A$. Note that the term $\dot \chi_{0}\delta \chi < 
O(\tau_{*}
{m\over g}\dot \phi_{0}\delta \phi)$ is not taken into account. It can
be said that the perturbation of the matter field $\delta \chi$ induces
the metric perturbation not directly, but through 
the excitation of the inflaton perturbation $\delta \phi$. We have:
$$A\approx -{\phi_{0}K\over \tau_{*}}e^{2\mu_{max}\tau}R(\tau), \eqno 37$$
where the function of time
$$R(\tau)=\lbrace {(-1)}^{j}s_{0}^{j}\sin \tau +{1\over 2}\sum_{l=0}^{l=j}
c_{l}(\tau -\tau_{l}+{\sin {2\tau} \over 2})\rbrace e^{-2\mu_{max}\tau},
\eqno 38$$
is constrained by the condition $|R(\tau)|< O(1)$.   
The coefficient $B$ is additionally damped with respect to $A$ in the long wave
limit, and is not calculated here. The vacuum 
expectation value for $A^{2}$ has a form:
$$\left <A^{2} \right >
\approx {4\pi \phi_{0}^{2}|K|^{2}|R(\tau)|^{2}\over \tau_{*}^{2}}e^{4\mu_{max}\tau_{*}}
\int k^{3} d\ln k \approx {2\over 3\pi^{2}}{g\over m}{|R(\tau)|^{2}\over 
\tau_{*}^{6}}m^{2}e^{-3N+4\mu_{max}\tau_{*}}\int e^{-3n}dn, \eqno 39$$
where $n=\ln {({a_{0}m\over k})}$ is e-folds number, corresponding to the wavenumber $k$,
and we assume that $\chi_{*}\sim {m\over g}e^{-3N/2}$.
Let us remind that for the modes with wavelengths of order of the present horizon scale
$n\approx N\approx 50$. The r.m.s value of the metric perturbation
$$A_{r.m.s.}(n)\approx \sqrt {{2\over 3\pi^{2}}}{({g\over m})}^{1/2}{|R(\tau)|\over \tau_{*}^{3}}m
e^{-3/2(N+n)+2\mu_{max}\tau_{*}}, \eqno 40$$
is even much smaller than the estimate $(2)$ due to very small factor $< 10^{-9}$ in the
front of the exponent in the eq. $(40)$.

Thus we conclude that in our case 
the large-scale metric perturbation induced due to the coupling between
the matter field perturbations $\delta \chi$ and the background part
of the field $\chi_{0}$ is absolutely negligible.

There is another source of unavoidable
 metric perturbations induced by fluctuations of the 
energy density and the 
pressure of the condensate of the $\chi$ ``particles''. These fluctuations
are non-linear, and therefore give rise to the metric 
perturbations even in the absence of the classical field $\chi_{0}$. 
The calculation of this effect
is even more complicated problem than the calculations in the linear theory, but it is possible
again to roughly estimate the characteristic order of magnitude. Similar to the linear case,
there is no hope to find some imprints of this effect at super-large scales, but 
it might play an important role right before the end of preheating, at the
horizon scale $\lambda_{*}=H^{-1}(t_{*})={3\over 2}\tau_{*}m^{-1}$.
The relatively 
large metric perturbations at this scale (say, with r.m.s amplitude 
$\sim 10^{-2}\div 10^{-1}$) might lead to the copious production of primordial black holes,
and modify the evolution of the Universe right after the end of the  preheating stage
\footnote{ When calculating the abundance of primordial black holes,
one should take into account that the metric perturbations are non-Gaussian 
in that case. This can lead to a modification of the standard estimates, based 
on assumption of Gaussian statistics for the perturbations.}.
An overproduction of the
primordial black holes might constrain
the parameters of our model.           

To characterize the energy density fluctuations at some comoving scale $\sigma$, 
we introduce coarse-grained energy density operator:
$$\hat \epsilon_{c.g}=\int d^{3}yG_{\sigma}(x-y)\hat \epsilon (y), \eqno 41$$
where 
$$\hat \epsilon ={1\over 2}({\dot {\hat \chi}}^{2}+g^{2}\phi_{0}^{2}{\hat \chi}^{2})$$
is the operator of the energy density of the $\chi$ field, and the Gaussian window function
$$G_{\sigma}(\vec r)={(2\pi)}^{-3/2}\sigma^{-3}e^{-{r^{2}\over 2\sigma^{2}}}.$$
Consider the relative standard deviation 
$$\delta_{\epsilon}(\sigma)=\sqrt{{\left <{\hat \epsilon_{c.g}}^{2}\right >
(\sigma)\over {\left <\hat \epsilon \right >}^{2}}-1}. 
\eqno 42$$
The calculation of this quantity can be easily done in two limiting cases. Let introduce
the wavenumber $k_{\sigma}=\sigma^{-1}$ corresponding to the scale $\sigma$, and also
the wavenumber 
$$k_{0}={k_{crit}\over {(2\mu^{eff} \tau )}^{1/6}}\
=\sqrt {{2\over \pi}}{q_{0}^{1/4}a_{0}m\over {(2\mu^{eff})}^{1/6}},$$
corresponding to the modes which have been amplified during all first stage of preheating:
$k_{0}\sim k_{crit}(\tau_{in})$, and the wavenumber corresponding to the horizon scale:
$$k_{h}(\tau)=a(\tau)H(\tau)={2\over 3}{a_{0}m\over \tau^{1/3}}.$$ Let also average the 
quantity $(42)$ over many oscillations of the field $\chi$.
Then, 
if $k_{0} < k_{\sigma} < k_{crit}$, the deviation $\delta_{\epsilon}$ is close to $1$:
$\delta_{\epsilon}\approx 1$. If $ k_{\sigma} < k_{0}$, we have
$$\delta_{\epsilon}^{2}\approx {3\over 2\sqrt 2}{({k_{\sigma}\over k_{0}})}^{3}. \eqno 43$$
The eq. $(43)$ can also be rewritten in terms of ratio $k_{\sigma}/k_{h}$:
$$\delta_{\epsilon}\approx D{({k_{\sigma}\over k_{h}})}^{3/2}, \eqno 44$$
where
$$D={\pi^{3/4}\over 3}{{(2\mu^{eff})}^{1/4}\over q_{0}^{3/8}\sqrt \tau}\approx 2\cdot 10^{-3}
{\mu_{0.14}^{1/4}\over q_{4}^{3/8}\sqrt \tau_{2}}, \eqno 45$$
and $\tau_{2}={\tau \over 10^{2}}$. For the perturbations with wavelengths smaller than
the horizon wavelength, the metric perturbation $A$ is related to the energy density 
perturbation via Poisson equation: ${\Delta A \over a^{2}}\approx -\delta \epsilon$. Assuming 
that $\delta \epsilon \sim \delta_{\epsilon}\left < \hat \epsilon \right >$, 
and also that at the end of the
first stage of preheating the energy density of $\chi$ ``particles'' influences
the expansion law: $H^{2}\sim \left <\hat \epsilon \right >$, we have:
$$A(k_{\sigma})\sim {({k_{h}(\tau_{*})\over k_{\sigma}})}^{2}\delta_{\epsilon}=
D{({k_{h}(\tau_{*})\over k_{\sigma}})}^{1/2}, \eqno 46$$
for $k_{h}(\tau_{*})< k_{\sigma} < k_{0}$. For super-horizon perturbations the dependence
of the perturbations on $k$ should be the same as a dependence of a source of the 
perturbations, and 
$$A(k_{\sigma})\sim \delta_{\epsilon}= D{({k_{\sigma}\over {k_{h}(\tau_{*})}})}^{3/2}.
\eqno 47$$  
Thus, the characteristic 
metric perturbation induced by the fluctuations of the energy density
takes its maximal value at the horizon scale, and this value is smaller than $\sim 10^{-3}$
if $q_{0} > 10^{4}$.

Similar to the case of linear perturbation, we did not take into account the contribution
of $\chi$ 'particles' in the dynamics of background model when calculating the amplitude of
the metric perturbations induced by the non-linear terms. However, this contribution can
change the parameters of the background model (such as, e.g. the expansion rate) only on
factor of order two, and we believe this modification cannot change significantly
our estimates
\footnote{Recently, this question as well as the generation of the linear perturbations
in the same model
has been considered numerically in the paper [11]. The authors came to the similar
conclusion that the perturbations of both types are very small at the large scales. 
I am very grateful to Dr. K. Jedamzik for drawing my attention to that paper.}.

\bigskip
{\large\bf Acknowledgments}
\bigskip

The author acknowledges Lev Kofman and especially
Vladimir Lukash for many valuable discussions, and also
Alexander Polnarev for useful comments. Queen Mary and Westfield College of London University is
acknowledged for the hospitality during the preparation of this paper in press. This work was
supported in part by the Danish Research Foundation through its establishment of the Theoretical
Astrophysics Center.

\end{document}